\documentclass[%
 reprint,
amsmath,amssymb,aps,
]{revtex4-1}

\usepackage{lipsum}
\usepackage{graphicx}
\usepackage{dcolumn}
\usepackage{bm}
\usepackage{natbib}

\usepackage{esint}
\usepackage{graphicx}
\usepackage{textcomp}
\usepackage{amsmath}
\usepackage{amsfonts}
\usepackage{mathtools}
\usepackage{color}
\usepackage{mathrsfs}
\usepackage{empheq}
\usepackage{tcolorbox}
\usepackage{fancyhdr}
\usepackage{datetime}
\usepackage{lipsum}
\usepackage{graphicx}
\usepackage{dcolumn}
\usepackage{bm}
\usepackage{color}
\usepackage{xspace}
\usepackage{amsmath}
\usepackage{nccmath}

\DeclareMathOperator{\Tr}{Tr}
\newcommand*\xb{\mathbf{x}}
\newcommand*\bq{\mathbf{Q}}
\newcommand*\bu{\mathbf{u}}

\newcommand*\bnu{\bm{\nu}}
\newcommand*\bpsi{\bm{\psi}}

\newcommand*\dxx{\text{d}^2\mathbf{x}~}

\newcommand{\fig}{FIG. }
\newcommand{\eqn}{EQN. }

\begin{document}


\title{Optimal Control of Active Nematics}

\author{Michael M. Norton}
\email{mike.m.norton@gmail.com}
\affiliation{Center for Neural Engineering,
Department of Engineering Science and Materials, Pennsylvania State University, University Park, Pennsylvania 16801}
\affiliation{
Physics Department, Brandeis University, Waltham, Massachusetts 02453
}
\author{Piyush Grover}
\affiliation{Mechanical and Materials Engineering, University of Nebraska - Lincoln, Lincoln Nebraska 68588}
\author{Michael F. Hagan}
\affiliation{
Physics Department, Brandeis University, Waltham, Massachusetts 02453
}
\author{Seth Fraden}
\email{fraden@brandeis.edu}
\affiliation{
Physics Department, Brandeis University, Waltham, Massachusetts 02453
}

\date{\today}

\begin{abstract}
In this work we present the first systematic framework to sculpt active nematics systems, using optimal control theory and a hydrodynamic model of active nematics. We demonstrate the use of two different control fields, (1) applied vorticity  and (2) activity strength, to shape the dynamics of an extensile active nematic that is confined to a disk. In the absence of control inputs, the system exhibits two attractors, clockwise and counterclockwise circulating states characterized by two co-rotating topological $+\frac{1}{2}$ defects. We specifically seek spatiotemporal inputs that switch the system from one attractor to the other; we also examine phase-shifting perturbations. We identify control inputs by optimizing a penalty functional with three contributions: total control effort, spatial gradients in the control, and deviations from the desired trajectory. This work demonstrates that optimal control theory can be used to calculate non-trivial inputs capable of restructuring active nematics in a manner that is economical, smooth, and rapid, and therefore will serve as a guide to experimental efforts to control active matter.
\end{abstract}

\pacs{Valid PACS appear here}

\maketitle

Active matter represents a broad class of materials and systems comprising interacting and energy-consuming constituents. Systems ranging from cytoskeletal proteins to bird flocks are unified by their common ability to spontaneously manifest collective behaviors on a scale larger than the individual agents. One of the promises of active matter research is that it will enable the design of new self-organizing materials that possess the life-like property of switching between distinct, robust, nontrivial dynamical states or configurations in response to external stimuli \cite{Needleman2017}. Towards rationally designing active materials with functional properties, we apply optimal control theory to an active nematic material, an important subclass of active matter that includes bacterial films and cell colonies \cite{Marchetti2013}, to switch the system between dynamical attractors in an optimally smooth, rapid and efficient manner. 

\begin{figure}
\includegraphics[width=\columnwidth]{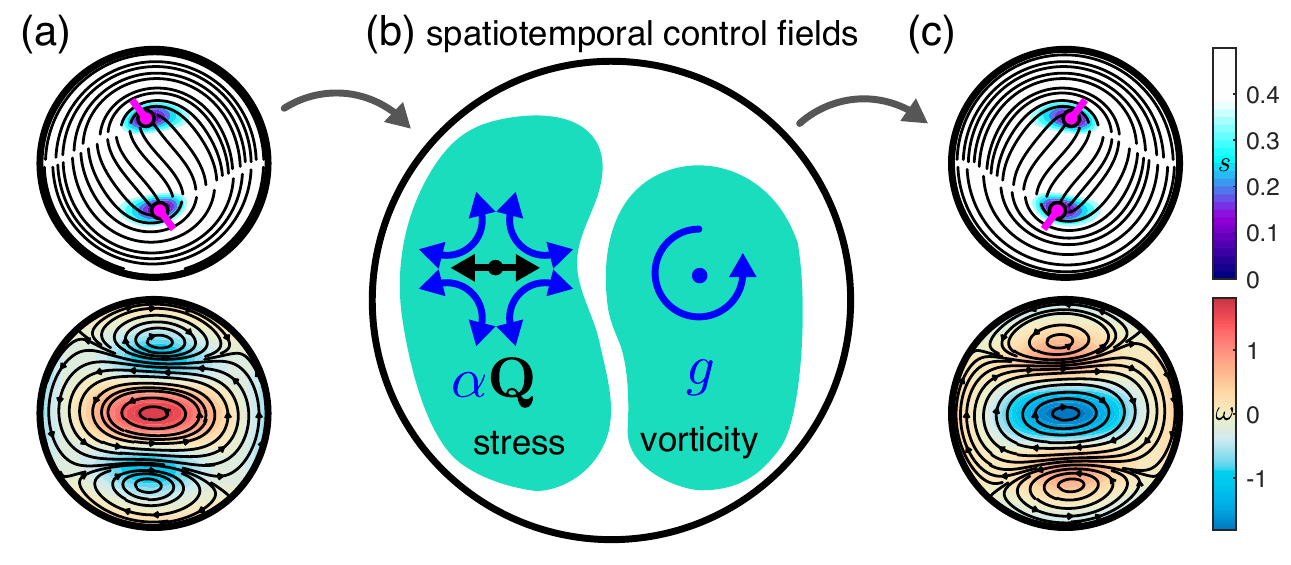}
\caption{(a) Director $\mathbf{n}$ with degree of order $s$ and defects in magenta (top) and flow field $\bu$ with vorticity $\omega$ (bottom), for initial condition in the counterclockwise circulating attractor. (b) Schematic showing two modes of control input. (c) Director and flow field of the clockwise target configuration.}\label{fig.chiral}
\end{figure}

Here we present a concrete paradigm for applying optimal control to active matter. This theoretical work is motivated by a model experimental active matter system comprising microtubules and motor proteins that utilizes ATP fuel to slide microtubules and thereby generate extensile stress\cite{Sanchez2012,Decamp2015, Lemma2020}. When compacted into a dense quasi-2D layer, these microtubules organize into a nematic with strong, local orientational order. Extensile stresses in active nematics drive instabilities that create motile topological defects and chaotic hydrodynamics \cite{Giomi2015, Doostmohammadi2016, Gao2017, Chen2018, Sokolov2019, MartinezPrat2019}. In order to harness the chemomechanical abilities of these materials to do useful work, these dynamics need to be controlled \cite{Gompper2020}. Experimentally, this has been accomplished through physical means by introducing anisotropy in the friction of the underlying surface using liquid crystals~\cite{guillamat2016control, Guillamat2017a}, and through confinement within hardwall boundaries ~\cite{Hardouin2019, Opathalage2019}. While these approaches can radically alter the dynamics of active nematics, by corralling defects into lanes or regular trajectories, the potential for spatio-temporal actuation with these methods is limited. Recently, light-activated motor complexes have been created, allowing active stress to be spatiotemporally modulated by an external light source in microtubule gels \cite{Ross2019} and nematics \cite{Zhang2019}. In these promising demonstrations, the control targets are relatively simple. This allows intuition, and trial and error, to inform suitable ad hoc control inputs. However, to systematically achieve more elaborate configuration goals, a framework is needed that includes a dynamical model of the system.

\begin{figure*}
\includegraphics[width=1\textwidth]{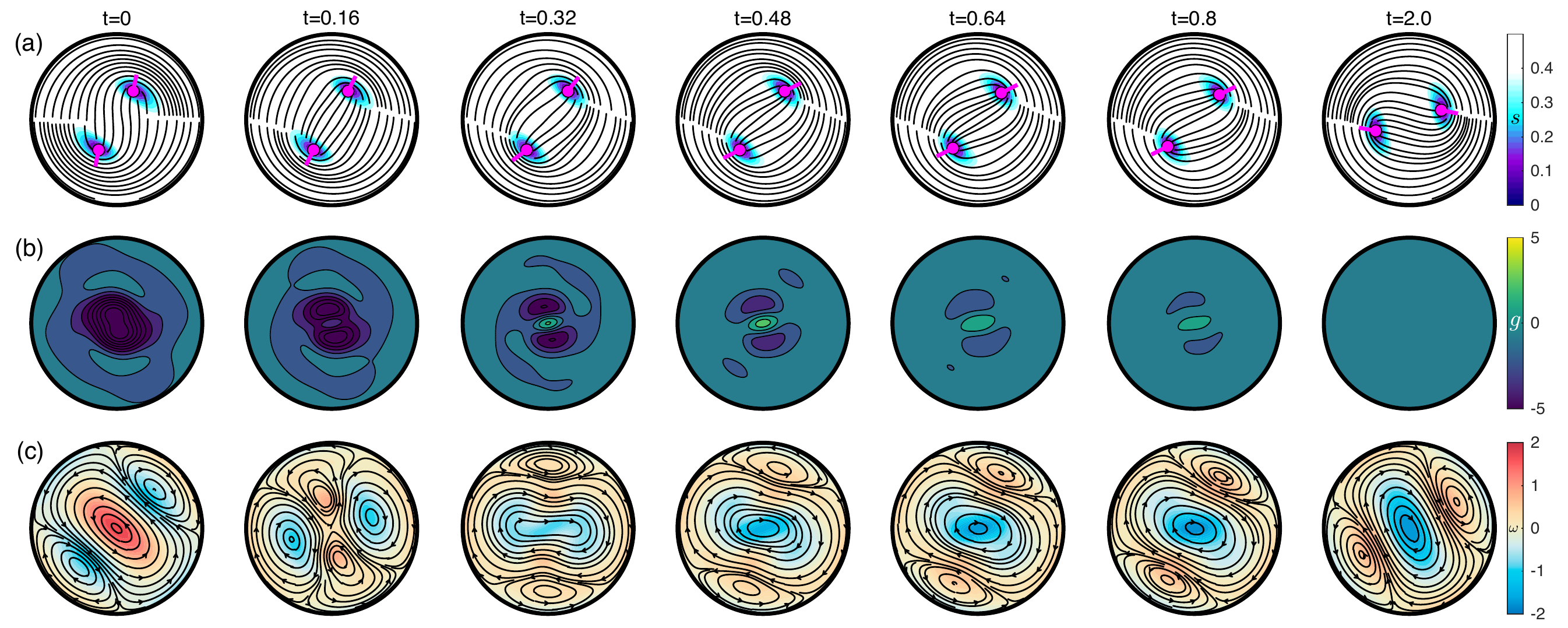}
\caption{Transformation from counterclockwise to clockwise rotation using applied vorticity actuation. (a) Director field and degree of order, (b) applied vorticity $g\left(\xb,t\right)$ and (c) velocity field and vorticity, with $W=600$, $t_f=2$. See movie S1.}\label{fig.ccw2cwvorticity}
\end{figure*}

To that end, we consider the problem of driving an active nematic between two attractors, by formally applying optimal control theory to a nemato-hydrodynamic model subject to constraints that promote functionality and solving for the necessary control inputs. Our control goal is therefore not to regulate or stabilize the system, but instead to steer the system between two distinct configurations in a manner that is optimally smooth, efficient and rapid. We consider separately two spatiotemporal control fields: (1) an applied rotation rate $g\left(\xb,t\right)$ that rotates the nematic director $\bq$ directly, and (2) the active stress strength $\alpha\left(\xb,t\right)$ which acts through the momentum equation and experimentally can be controlled through light input \cite{Ross2019,Zhang2019}. The former does not currently have an experimental analog but nothing in principle prevents engineering such a control field.

\begin{figure*}
\includegraphics[width=1\textwidth]{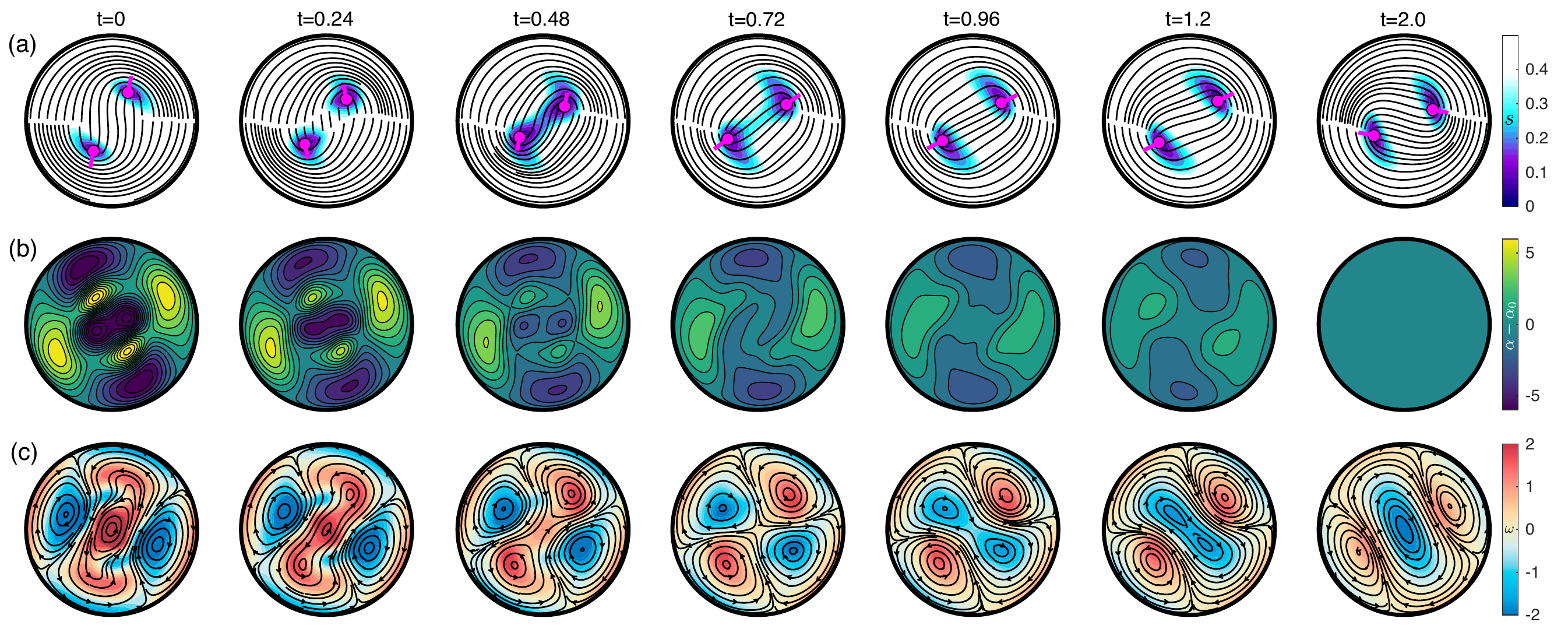}
\caption{Transformation from counterclockwise to clockwise rotation using active stress actuation, (a) director field and degree of order, (b) active stress $\alpha\left(\xb,t\right)-\alpha_0$ and (c) velocity field and vorticity, $W=900$, $t_f=2$. See movie S2.}\label{fig.ccw2cwstress}
\end{figure*}

We consider an active nematic already corralled by confinement in a disk with strong parallel anchoring and no-slip boundary conditions, such that it does not exhibit chaotic dynamics \cite{Norton2018a}. Instead, the system produces two stable limit cycle attractors characterized by two, motile $+\frac{1}{2}$ disclinations that perpetually orbit the domain at fixed radius in either the clockwise or counterclockwise direction, note the handedness of the defect configurations in \fig\ref{fig.chiral}\emph{a} and \emph{c}. I.e., these attractors are mirror images of each other, a consequence of the system dynamics' equivariance under reflection. We emphasize that while these are steady states of the system, they are maintained by a constant flux of energy through the extensile active stress and are therefore minimal self-organized attractors of the active nematic material. In the absence of activity, the director field would relax to a motionless equilibrium configuration. As an exemplar application of optimal control theory, we have identified the spatio-temporal actuation of either applied vorticity or active stress that re-arranges the nematic director field and moves the system from one attractor to the other, while optimally balancing the amount of control input against the penalty for deviations from the target configuration (\fig\ref{fig.ccw2cwvorticity} and \ref{fig.ccw2cwstress}, movies S1 and S2). This is akin to the process of gait switching in neuroscience. It has been shown that even small networks of oscillators are capable of multiple rhythms that are accessible through external inputs \cite{wojcik2014key, Lodi2019}. We also consider phase-shifting perturbations within one attractor using both control fields, movies S3 and S4.



\begin{figure}
\includegraphics[width=1\columnwidth]{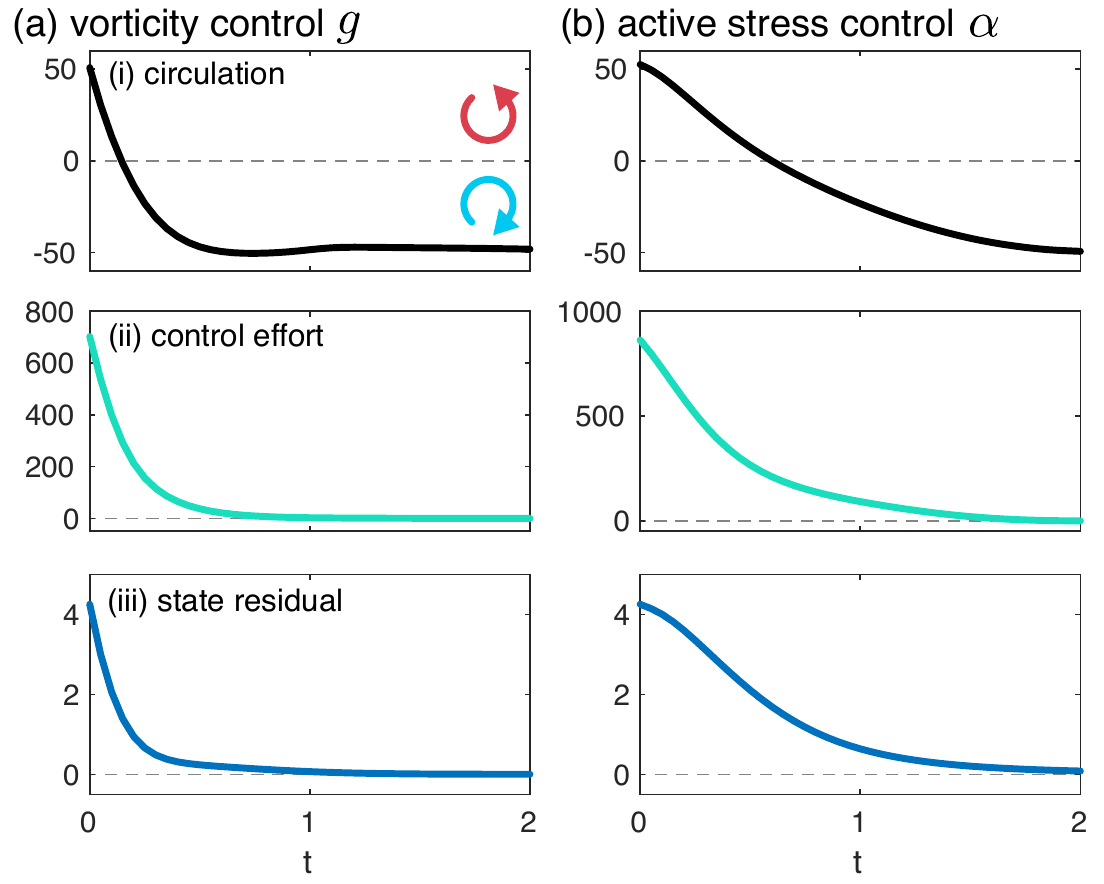}
\caption{ Evolution of bulk system properties during counterclockwise to clockwise maneuvers using (a) vorticity control $g\left(\xb,t\right)$, \fig\ref{fig.ccw2cwvorticity}, and (b) active stress strength $\alpha\left(\xb,t\right)$, \fig\ref{fig.ccw2cwstress}: (i) circulation $\int_{\Omega}{\dxx}\bu\cdot\hat{\mathbf{e}}_{\theta}$, control effort for (a) $\frac{1}{2}\int_{\Omega}{\dxx}g^2$ or (b) $\frac{1}{2}\int_{\Omega}{\dxx}\left(\alpha-\alpha_0\right)^2$, and (iii) residual between system state and target $\frac{1}{4}\int_{\Omega}{\dxx}\Delta\bq:\Delta\bq$.}\label{fig.circulation}
\end{figure}

\textbf{Nemato-Hydrodynamic Model.} To develop our optimal control problem, we utilize a previously explored continuum nematohydrodynamic model~\cite{BerisEdwardsThermo, Norton2018a}, which we restate here in dimensionless form. We choose this model for simplicity but note that other models can be put into the same control framework \cite{Thampi2016b, Gao2017}. As a minimal representation of our system, we use a single-fluid model whose state is described by the dimensionless nematic order tensor $\mathbf{Q}=s\rho\left[\mathbf{n}\otimes\mathbf{n}-\left(1/2\right)\mathbf{I}\right]$ and fluid flow field $\mathbf{u}$. $\mathbf{Q}$ describes both the local orientation $\mathbf{n}$ and degree of order $s$ of the nematic, and is scaled by the nematic density $\rho$ such that ($\rho s = \sqrt{2 \Tr \mathbf{Q}^2}$). The coupled dynamics are given by
\begin{multline}
 \partial_{t}\mathbf{Q}+\nabla\cdot\left(\mathbf{u}\mathbf{Q}\right)-\left(\left(\mathbf{\Omega+G}\right) \mathbf{Q}-\mathbf{Q}\left(\mathbf{\Omega+G}\right)\right)
-\lambda\mathbf{E}-\mathbf{H}\\
=f_{\bq}\left(\bq,\bu,g\right)=0
\label{eqn.Qdynamics}
\end{multline}
Along the boundary $\left.\mathbf{Q}\right|_{\partial\Omega}=s^{*}\rho \left[\mathbf{t}\otimes\mathbf{t}-\left(1/2\right)\mathbf{I}\right]$, with boundary tangent $\mathbf{t}$ and degree of order $s^*=\sqrt{2}$ associated with a fully ordered nematic in the limit $\rho\rightarrow\infty$ \cite{Putzig2015}. Kinematic terms and free-energy relaxation both contribute to the dynamics of $\mathbf{Q}$. The kinematic terms depend on the local fluid flow velocity and gradients, with $\Omega_{ij}=\frac{1}{2}\left(\partial_{j}u_{i}-\partial_{i}u_{j}\right)$ as the antisymmetric vorticity tensor and $E_{ij}=\frac{1}{2}\left(\partial_{j}u_{i}+\partial_{i}u_{j}\right)$ as the symmetric strain rate tensor. The vorticity tensor is augmented by an applied field $\mathbf{G}=\frac{1}{2}\begin{psmallmatrix}0 & -g\\ g & 0\end{psmallmatrix}$, where $g$ is one of the control inputs we consider in this paper, applied vorticity. The relaxational terms $\mathbf{H}$ are proportional to variations of the system free energy, $\mathbf{H}=-\left(\beta_1-\beta_2 \bq:\bq\right)\bq-2\nabla^2\bq$.
Momentum conservation in the Stokes limit, incompressibility constraint $\nabla\cdot\mathbf{u}=0$, and boundary conditions $\left.\bu\right|_{\partial\Omega}=0$ govern the fluid flow
\begin{equation}
\eta\nabla^{2}\mathbf{u}-\nabla P-\nabla\cdot\left(\alpha \mathbf{Q}\right)=f_{\bu}\left(\bq,\bu,\alpha\right)=0,
\label{eqn.udynamics}
\end{equation}
with pressure $P$ and strength of activity $\alpha$. The active stress, $-\alpha\mathbf{Q}$ corresponds to an extensile dipole force density \cite{Simha2002, Marchetti2013, Thampi2014}. The scaling factor $\alpha$ serves as the second form of spatiotemporal control input that we consider.


\textbf{Optimal Control.} We seek a spatio-temporal input field, either $g\left(\xb,t\right)$ or $\alpha\left(\xb,t\right)$, that drives the system towards a desired director field configuration $\bq^*$ by minimizing the following scalar cost functional $\mathscr{J}$
\begin{align}
&\mathscr{J} =
\frac{1}{2}\int_{0}^{t_f}{\text{d}t}\int_{\Omega}{\dxx}
\left[g^2\right.
+\Gamma_g\nabla g\cdot\nabla g \label{eqn.costfunction}\\
&+\left(\alpha-\alpha_0\right)^2
+\Gamma_{\alpha}\nabla\alpha\cdot\nabla\alpha+W\frac{1}{2}\left.\Delta\bq:\Delta\bq\right],\nonumber
\end{align}
subject to \eqn\ref{eqn.Qdynamics} and \eqn\ref{eqn.udynamics}. We pose the control problem as a tracking problem by quadratically penalizing the deviations $\Delta\bq=\bq\left(\xb,t\right)-\bq^*\left(\xb,\theta+t-t_f\right)$ from the desired state throughout the control window $t\in \left[0,t_f \right]$. $\bq^*$ is selected from a pre-calculated time-periodic solution at phase $\theta$, \fig\ref{fig.chiral}.   
Control actuation is also penalized quadratically with either $\left(\alpha-\alpha_0\right)^2$ or $g^2$. We penalize deviations from $\alpha_0$ rather than $\alpha$ itself to maintain the intrinsic dynamics of the material as much as possible. Finally, we promote smoothness on $\alpha$ and $g$ by additionally penalizing $\nabla \alpha\cdot\nabla \alpha$ and $\nabla g \cdot\nabla g$, with weights $\Gamma_\alpha$ and $\Gamma_g$. This is more crucial when $\alpha$ is the control input, since gradients in $\alpha$ can be exploited to achieve arbitrarily large forces, which we want to discourage. Including these three penalties creates a control problem with opposing forces; the solutions we identify optimally balance matching the desired trajectory quickly against applying inputs to the system. We can sacrifice accuracy but use less control by decreasing the weight $W$ on the state penalty, or alternatively arrive more rapidly at the target configuration by increasing $W$. We note that changing the norm on the control inputs can be used to promote spatio-temporally sparse (localized) actuation, in contrast to the smooth and distributed control inputs we identify \cite{Ryll2016}.

Following Pontryagin's theorem \cite{Kirk1970,Lenhart}, we constrain our search of optimal state trajectories to those that obey the system dynamics by introducing Lagrange multipliers $\bpsi\left(\xb,t\right)\in\mathbb{R}^{2\times 2}$, $\bnu\left(\xb,t\right)\in\mathbb{R}^2$, and $\phi\left(\xb,t\right)\in\mathbb{R}^1$, which are the adjoint or costate variables for $\bq$, $\bu$, and $P$, respectively, and augmenting the original cost function \eqn\ref{eqn.costfunction} to give  $\mathscr{L}=\mathscr{J}+\int_{0}^{t_f}{\text{d}t}\int_{\Omega}{\dxx}\left[\bnu\cdot\mathbf{f}_\bu+\bpsi:\mathbf{f}_{\bq}+\phi\left(\nabla\cdot\bu\right)\right]$. The conditions for optimality are $\frac{\delta \mathscr{L}}{\delta \bpsi}$, $
\frac{\delta \mathscr{L}}{\delta \bnu}$, $
\frac{\delta \mathscr{L}}{\delta \phi}$, $
\frac{\delta \mathscr{L}}{\delta \mathbf{Q}}$, $\frac{\delta \mathscr{L}}{\delta \mathbf{u}}$,
$\frac{\delta \mathscr{L}}{\delta P}$,
$\frac{\delta \mathscr{L}}{\delta \alpha}$,
$\frac{\delta \mathscr{L}}{\delta g}=0$. The first three conditions simply return the original nematohydrodynamic equations governing $\{\bq,\bu,P\}$, \eqn\ref{eqn.Qdynamics} and \eqn\ref{eqn.udynamics}. The following three conditions yield the dynamical equations for the adjoint variables $\{\bpsi,\bnu,\phi\}$:

\begin{align}
&-\nabla^2\bnu-\nabla\phi+\mathbf{h}_1=0, \nabla\cdot\bnu=0\label{eqn.nudynamics}\\
&W\left(\bq-\bq^*\right)-\partial_t\bpsi-\bu\cdot\nabla\bpsi-2\nabla^2\bpsi+\alpha \mathbf{h}_2+\mathbf{h}_3\nonumber\\
&+2\bq\beta_2\left(\bq:\bpsi\right)-\bpsi\left(\beta_1-\beta_2\bq:\bq\right)=0,
\label{eqn.psidynamics}
\end{align}
with final conditions $\{\bnu,\bpsi\}\left(\xb,t_f\right)=0$ and boundary conditions $\left.\{\bnu,\bpsi\}\right|_{\partial\Omega}=0$. Full expressions for the terms $\mathbf{h}_{1-3}$ are stated here
\footnote{$h_{1x}=\psi_{xx}\partial_x Q_{xx}+\psi_{xy}\partial_x Q_{xy}+\partial_y\left(Q_{xy}\psi_{xx}-Q_{xx}\psi_{xy}\right)+\lambda\left(\partial_x\psi_{xx}+\partial_y\psi_{xy}/2\right)$,
$h_{1y}=\psi_{xx}\partial_y Q_{xx}+\psi_{xy}\partial_y Q_{xy}+\partial_x\left(Q_{xx}\psi_{xy}-Q_{xy}\psi_{xx}\right)+\lambda \partial_x\psi_{xy}/2$,
$h_{2x}=\partial_y\nu_y-\partial_x\nu_x$,
$h_{2y}=-\left(\partial_y \nu_{x}+\partial_x \nu_y\right)$,
$h_{3x}=\psi_{xy}\left(\partial_y u_x-\partial_x u_y\right)$,
$h_{3y}=\psi_{xx}\left(\partial_x u_y-\partial_y u_x\right)$}, where we've used $Q_{xy},\psi_{xy}=Q_{yx},\psi_{yx}$ and $Q_{xx},\psi_{xx}=-Q_{yy},\psi_{yy}$.
The final two optimality conditions $\frac{\delta \mathscr{L}}{\delta \alpha},
\frac{\delta \mathscr{L}}{\delta g}=0$ constrain the control inputs:
\begin{align}
&\left(\alpha-\alpha_0\right)-\Gamma_{\alpha}\nabla^2\alpha
\label{eqn.optimalcond_control_stress}
\\
&-\left(Q_{xx}\left(\partial_x\nu_x-\partial_y\nu_y\right)+Q_{xy}\left(\partial_y\nu_x+\partial_x\nu_y\right)\right)=0
\nonumber
\\
&g-\Gamma_g\nabla^2 g+\left(Q_{xy}\psi_{xx}-Q_{xx}\psi_{xy}\right)=0
\label{eqn.optimalcond_control_torque}
\end{align}


We solve the coupled PDE-system and constraints using the direct-adjoint-looping (DAL) method \cite{Kerswell2014}. We consecutively solve the forward dynamics \eqn\ref{eqn.Qdynamics} and \eqn\ref{eqn.udynamics}, and adjoint dynamics \eqn\ref{eqn.psidynamics} and \eqn\ref{eqn.nudynamics}. The latter are solved backwards in time and thus are responsible for propagating the residuals of the director field $\Delta\bq$. After each backward run, the control fields are updated via gradient descent using the \eqn \ref{eqn.optimalcond_control_torque} for applied vorticity or \eqn \ref{eqn.optimalcond_control_stress} for stress. Gradient descent step sizes are chosen using the Armijo backtracking method \cite{Borzi2011}. The process is repeated until the cost function \eqn\ref{eqn.costfunction} converges to a desired tolerance $\sim 10^{-4}$. The first forward step is completed using a naive initial guess of $\alpha\left(\mathbf{x},t\right)=\alpha_0$ for the stress-actuated case or $g=-0.25$ for the vorticity-actuated case.
For all computations $\lambda=1$ and $\Gamma_{g,\alpha}=0.1$. We restrict ourselves to a domain size of dimensionless radius $R=6.5$ and a dimensionless baseline active stress $\alpha_0=5$ that produces a stable periodic solution consisting of a single fluid vortex driven by two $+\frac{1}{2}$ defects (\fig\ref{fig.chiral}) \cite{Norton2018a}. Both clockwise and counterclockwise circulating states are pre-calculated and used as $\bq^*$. The state and adjoint fields are integrated using the finite element analysis software COMSOL.


\textbf{Results.} We first consider chirality switching using applied vorticity $g$, see \fig\ref{fig.ccw2cwvorticity} and movie S1. At $t=0$ the applied control field (second row) is strongest in the center of the disk and opposes the native vorticity (last row). The amplitude of the control quickly fades as the director field (first row) transitions through a symmetric, dipolar configuration ($t=0.16$) to the other attractor. After the defects begin circulating in the desired clockwise direction, small amounts of control are applied to adjust the director field so that it matches the target trajectory (movies S1-4 all show both the solution $\bq$ and target $\bq^*$ for reference). We next consider the same control goal using the active stress strength $\alpha$ as the control input, \fig\ref{fig.ccw2cwstress}. In this case a combination of strong extensile and weak contractile stresses are used to again pull the defects towards the symmetric dipolar configuration ($t=0.96$) and then nudge them towards the other attractor. \fig\ref{fig.circulation} summarizes the dynamics of the system for both vorticity and stress control by plotting the evolution of three spatially-integrated quantities: (i) circulation, which coarsely describes the proximity of the system to one of the circulating attractors (note that the zero crossing aligns with the time at which the system transitions through the symmetric, dipolar configuration), (ii) control effort, and (iii) residual of the director which measures the approach of the system $\bq$ to the target trajectory $\bq^*$.

We next consider phase-shifting maneuvers actuated by vorticity (movie S3) and stress (movie S4). In the former, additional positive vorticity is applied to the interior of the disk, between the two $+\frac{1}{2}$ defects, while negative vorticity is applied in two regions at larger radii. These regions of opposing applied vorticity act on either side of the defects like enmeshed gears, to rapidly advance the position of the defects and thus the phase. This action adds noticeable twist to the director ($t=0.24$) that relaxes once the control fades. In the latter, additional active stresses are added at large radii, stopping circulation and temporarily moving the director to a symmetric, dipolar configuration ($t=0.68$), similar to the action seen in the attractor switching maneuver explored above in \fig\ref{fig.ccw2cwstress} and movie S2. Active stresses then break the symmetry to resume circulation at the desired phase.

For both chirality switching and phase shifting, given the periodic nature of the starting and target configurations, it is likely that multiple solutions exist. For example, for every phase-advancing solution, there is likely a phase-delaying solution that achieves the same goal. This situation becomes more complex when switching between attractors and the phases of each attractor come into play. Thus, while we have used an optimal control framework, we cannot guarantee that the solutions found are globally optimal when multiple solutions exist; in this sense the control framework provides a means to generate \emph{a} physically-informed and plausible control solution. We speculate that isochron and isostable reduction of the full order control problem may yield physical insights and permit exploring the range of control scenarios more effectively \cite{Wilson2015a, Monga2018, Wilson2019a}. Creating a more parsimonious model would also reduce computational time, easing experimental implementation.

\textbf{Conclusion.} A grand challenge in active matter is to develop systems built from simple building blocks that manifest pre-programmed spatiotemporal dynamics \cite{Gompper2020}. The two circulating states we explore are emergent dynamical attractors that spontaneously self-assemble from an active liquid crystal when parameters are tuned correctly. This work paradigmatically demonstrates that one attractor can be effectively \emph{re-assembled} into a \emph{new} attractor through the proper choice of system inputs.


While the active nature of the fluid we consider here gives rise to dynamics fundamentally different than those of driven, passive fluids, connections can be made between our work and control of classical turbulence. The attractors we explore are examples of exact coherent structures (ECSs), which are typically exact solutions to the Navier-Stokes equations. The dynamics of a turbulent flow can be thought of as a meandering path that visits multiple ECSs through heteroclinic connections \cite{Suri2017,Suri2018}. While we have applied perturbations to switch basins of two attractors, one could envision a similar class of control problems that aim to stabilize certain ECSs or remove connecting orbits between ECSs to delay transition to the chaotic or (mesoscale) turbulent regime in active nematic systems \cite{Doostmohammadi2016b, Opathalage2019}. Further, while we explored spatio-temporally smooth control inputs, changing the norm on the control penalty can be used to promote sparsity, resulting in control inputs that are spatially or temporally localized \cite{Ryll2016}.

In addition to artificial active systems, living systems also present a potential application of control theory \cite{Lenhart}. For example, stress gradients radically rearrange cells during embryonic development  \cite{Streichan2018} and wound healing \cite{Cohen2014, Turiv2020, Zajdel2020}. One can use the inverse problem framework of optimal control to design the stress fields necessary to achieve a particular morphological change, thereby guiding both experimental and medical device design.


\begin{acknowledgments}
\textbf{Acknowledgements.} This work was supported by the NSF MRSEC-1420382, NSF-DMR-2011486, NSF-DMR1810077, and DMR-1855914 (MMN, MFH and SF). PG was supported by UNL startup fund. Computational resources were provided by the NSF through XSEDE computing resources (MCB090163) and the Brandeis HPCC which is partially supported by NSF-DMR-2011486. We thank Aparna Baskaran and Chaitanya Joshi for their helpful discussions.
\end{acknowledgments}

\bibliographystyle{apsrev4-1}



%

\end{document}